\pdfoutput=1
\documentclass[final]{aipproc}
\layoutstyle{6x9}

\usepackage{amsmath}
\usepackage{amssymb}
\usepackage{mathtools}
\usepackage{tikz}
\usetikzlibrary{arrows,shapes,decorations,automata,backgrounds,petri, positioning}
\usepackage{cancel}

\DeclareMathOperator{\Prob}{\operatorname{Pr}}

\newcommand{\Cond}[2]{\Prob\left({#1}\middle|{#2}\right)}

\newcommand{\SerAmp}[1]{#1}

\newcommand{\Product}[2]{{#1}\cdot{#2}}
\newcommand{\Sum}[2]{{#1}+{#2}}
\newcommand{\Cc}[1]{\overline{#1}}
\newcommand{\abs}[1]{\left|{#1}\right|}
\newcommand{\C}{\mathbb{C}}

\newcommand{\R}{\mathbb{R}}

\newcommand{\f}{\operatorname{\textbf{f}}}

\newcommand{\AmpEl}[3]{{\SerAmp{#1}}_{{#2}{#3}}}
\newcommand{\Hop}{\operatorname{\textbf{H}}}
\newcommand{\Gop}{\operatorname{\textbf{G}}}
\newcommand{\Htwo}[4]{\Hop\negmedspace\left[\begin{matrix}#1&\!\!#2\\#3&\!\!#4\end{matrix}\right]}
\newcommand{\Gtwo}[4]{\Gop\negmedspace\left[\begin{matrix}#1\!\!\!&;&\!\!\!#2\\ #3\!\!\!&;&\!\!\!#4\end{matrix}\right]}
\newcommand{\Htwosmall}[4]{\Hop\negmedspace\left[\begin{smallmatrix}#1&\!\!#2\\#3&\!\!#4\end{smallmatrix}\right]}

\newcommand{\Hthree}[9]{\Hop\negmedspace\left[\begin{matrix}#1&\!\!#2&\!\!#3\\#4&\!\!#5&\!\!#6\\#7&\!\!#8&\!\!#9\end{matrix}\right]}
\newcommand{\Hthreesmall}[9]{\Hop\negmedspace\left[\begin{smallmatrix}#1&\!\!#2&\!\!#3\\#4&\!\!#5&\!\!#6\\#7&\!\!#8&\!\!#9\end{smallmatrix}\right]}

\newcommand{\TikZstartFeynman}{\begin{tikzpicture}[scale=2,node distance=3cm,>=stealth',bend angle=45,auto]%
  \tikzstyle{state} = [circle,thick,fill=none,draw=black,minimum size=7mm]%
  \tikzstyle{nostate} = [circle,thick,fill=none,draw=none,minimum size=7mm]%
\tikzstyle{elstate} = [rectangle,thick,fill=none,draw=black,minimum size=7mm,rounded corners]%
  \tikzstyle{astate} = [circle,fill=none,draw=black,minimum size=7mm]%
  \tikzstyle{bstate} = [circle,double,fill=none,draw=black,minimum size=7mm]%
  \tikzstyle{every label}=[black]}
\newcommand{\TikZstartFeynmanSmall}{\begin{tikzpicture}[scale=2,node distance=1.7cm,>=stealth',bend angle=45,auto]%
  \tikzstyle{state} = [circle,thick,fill=none,draw=black,minimum size=6.5mm]%
  \tikzstyle{nostate} = [circle,thick,fill=none,draw=none,minimum size=6.5mm]%
 \tikzstyle{elstate} = [rectangle,thick,fill=none,draw=black,minimum size=6.5mm,rounded corners]%
  \tikzstyle{astate} = [circle,fill=none,draw=black,minimum size=6.5mm]%
  \tikzstyle{bstate} = [circle,double,fill=none,draw=black,minimum size=6.5mm]%
  \tikzstyle{every label}=[black]}
\newcommand{\TikZstartFeynmanMedium}{\begin{tikzpicture}[scale=2,node distance=2cm,>=stealth',bend angle=45,auto]%
  \tikzstyle{state} = [circle,thick,fill=none,draw=black,minimum size=6.7mm]%
  \tikzstyle{nostate} = [circle,thick,fill=none,draw=none,minimum size=6.7mm]%
 \tikzstyle{elstate} = [rectangle,thick,fill=none,draw=black,minimum size=6.7mm,rounded corners]%
  \tikzstyle{astate} = [circle,fill=none,draw=black,minimum size=6.7mm]%
  \tikzstyle{bstate} = [circle,double,fill=none,draw=black,minimum size=6.7mm]%
  \tikzstyle{every label}=[black]}
\newcommand{\TikZstartFeynmanReg}{\begin{tikzpicture}[scale=2,node distance=2.8cm,>=stealth',bend angle=45,auto]%
  \tikzstyle{state} = [circle,thick,fill=none,draw=black,minimum size=6.7mm]%
  \tikzstyle{nostate} = [circle,thick,fill=none,draw=none,minimum size=6.7mm]%
 \tikzstyle{elstate} = [rectangle,thick,fill=none,draw=black,minimum size=6.7mm,rounded corners]%
  \tikzstyle{astate} = [circle,fill=none,draw=black,minimum size=6.7mm]%
  \tikzstyle{bstate} = [circle,double,fill=none,draw=black,minimum size=6.7mm]%
  \tikzstyle{every label}=[black]}
\newcommand{\TikZend}{\end{tikzpicture} }
\begin{document}

\title{On the Origin of the Quantum Rules for Identical Particles}

\classification{\texttt{03.65.Ca}, \texttt{03.65.Ta} }

\author{Klil H. Neori}{
  address={University at Albany, SUNY}
}

\author{Philip Goyal}{
  address={University at Albany, SUNY}
}

\begin{abstract}
We present a proof of the Symmetrization Postulate for the special case of non-interacting, identical particles.  The proof is given in the context of the Feynman formalism of Quantum Mechanics, and builds upon the work of Goyal,~Knuth~and~Skilling~\cite{GoyalKnuthSkilling2010}, which shows how to derive Feynman's rules from operational assumptions concerning experiments.

Our proof is inspired by an attempt to derive this result due to Tikochinsky~\cite{Tikochinsky1988}, but substantially improves upon his argument, by clarifying the nature of the subject matter, by improving notation, and by avoiding strong, abstract assumptions such as analyticity.
\end{abstract}

\maketitle
\section{Feynman's Rules}

\begin{figure}[ht]
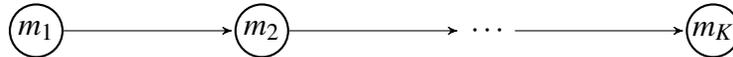

\centering
\TikZstartFeynman
  \begin{scope}
	\node [state] (m11) [label=center:{\(m_1\)}] {};
	\node [state] (m21) [right of=m11,label=center:{\(m_2\)}] {}
	edge 	[pre] (m11);
	\node [nostate] (m31) [right of=m21,label=center:{\(\cdots\)}] {}
	edge [pre] (m21);
	\node [state] (m41) [right of=m31,label=center:{\(m_K\)}] {}
	edge [pre] (m31);
  \end{scope}
\TikZend
  \caption{\emph{Sequence of Outcomes:} corresponding to the series of measurements~\(M_{(1)}, M_{(2)},\dots,M_{(K)}\).}
  \label{fig:seriesofresults}
\end{figure}

We begin by reviewing Feynman's rules. Suppose that we perform series of measurements: \(M_{(1)}, M_{(2)},\dots,M_{(K)}\), where a measurement \(M\) can have any number of results. For example, this could correspond to Stern-Gerlach experiments, or to multiple-slit measurements. These measurements are performed in series, leading to a corresponding sequence of outcomes: \([m_1, m_2,\dots,m_K]\). Graphically, we would represent it as in Fig.~\ref{fig:seriesofresults}.

\begin{figure}[ht]
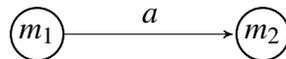
 
\centering
\TikZstartFeynman

  \begin{scope}
	\node [state] (m11) [label=center:{\(m_1\)}] {};
	\node [state] (m21) [right of=m11,label=center:{\(m_2\)}] {}
	edge 	[pre] node[above] {\(\SerAmp{a}\)} (m11);
	
  \end{scope}

\TikZend
  \caption{\emph{Amplitude for a Pair of Measurements:} the probability of this process, that is, of \(m_2\) occurring in the second measurement given that \(m_1\) had occurred in the first,  is: \(\Cond{m_2}{m_1}=\abs{\SerAmp{a}}^2\).}
  \label{fig:twomeasurementamp}
\end{figure}

Now, we wish to make predictions about these measurements. That is to say, given that \(m_1\) was measured in \(M_{(1)}\), what is the probability of the outcomes of the rest of the sequence being \(m_2,\dots,m_K\)? In the Feynman formalism, we attach an amplitude, to each outcome sequence, that is, a complex number within the unit circle. This amplitude incorporates the physical process that is involved in this series of measurements, and the probability for this particular result is the modulus of the amplitude squared. (See Fig.~\ref{fig:twomeasurementamp})

This association of an amplitude to each sequence of measurement outcomes is a representation of the set of sequences. One can apply operators to this set, which this representation should mirror through operations on the corresponding amplitudes.

\begin{figure}[ht]
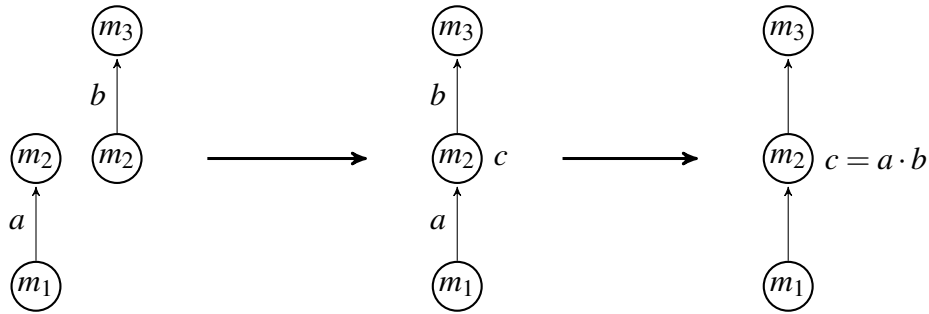
 
\centering
\TikZstartFeynmanSmall

\begin{scope} 
	\node [state] (m11) [label=center:{\(m_1\)}] {};
	\node [state] (m21) [above of=m11,label=center:{\(m_2\)}] {}
	edge 	[pre] node[left] {\(\SerAmp{a}\)} (m11);
	\node [nostate] (m31) [above of=m21] {};

	\node [nostate] (m12) [right=0.4cm of m11] {};
	\node [state] (m22) [above of=m12,label=center:{\(m_2\)}] {};
	\node [state] (m32) [above of=m22,label=center:{\(m_3\)}] {} 
	edge 	[pre] node[left] {\(\SerAmp{b}\)} (m22);
  \end{scope}  

  \begin{scope}[xshift=2.8cm] 

	\node [state] (m1) [label=center:{\(m_1\)}] {};
	\node [state] (m2) [above of=m1,label=right:\(\SerAmp{c}\),label=center:{\(m_2\)}] {}
	edge 	[pre]	node[left] {\(\SerAmp{a}\)} (m1);
	\node [state] (m3) [above of=m2,label=center:{\(m_3\)}] {}
	edge 	[pre] 	node[left] {\(\SerAmp{b}\)} (m2);

\end{scope}

\begin{scope}[xshift=5cm] 

	\node [state] (m'1) [label=center:{\(m_1\)}] {};
	\node [state] (m'2) [above of=m'1,label=right:{\(\SerAmp{c}=\Product{\SerAmp{a}}{\SerAmp{b}}\)},label=center:{\(m_2\)}] {}
	edge 	[pre] (m'1);
	\node [state] (m'3) [above of=m'2,label=center:{\(m_3\)}] {}
	edge 	[pre]  (m'2);

\end{scope}

 \draw [->,thick,line width=.4mm]
    ([xshift=0.6cm]m22 -| m22) -- ([xshift=-0.6cm]m2 -| m2);

 \draw [->,thick,line width=.4mm]
    ([xshift=0.7cm]m2 -| m2) -- ([xshift=-0.6cm]m'2 -| m'2);

\TikZend
  \caption{\emph{Product Rule}: \(\SerAmp{a}\) and \(\SerAmp{b}\) are the amplitudes of the two concatenated sequences. The amplitude of their concatenation is \(\SerAmp{c}=\Product{\SerAmp{a}}{\SerAmp{b}}\).}
  \label{fig:concat}
\end{figure}

One operation we can perform on sequences of outcomes is concatenation: given two sequences, the final outcome of the first being identical to the initial outcome of the second, we can rewrite them as a single sequence. Its amplitude is the product of those for the old. This is the \emph{Product Rule} (see Fig.~\ref{fig:concat}).

\begin{figure}[!ht]
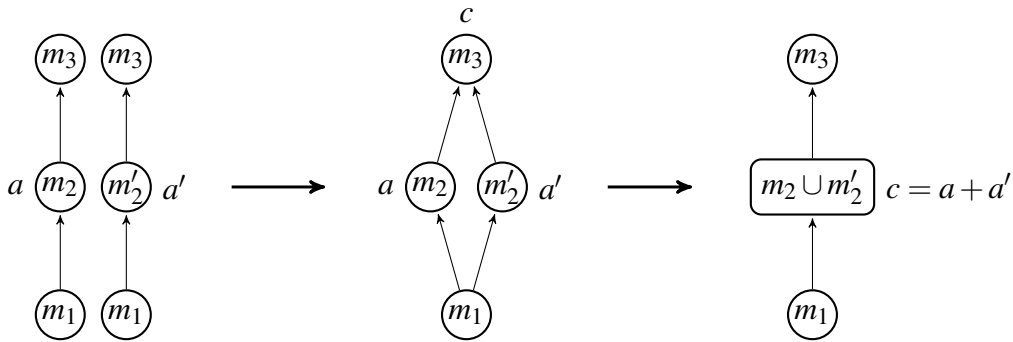
 
\centering
\TikZstartFeynmanSmall

  \begin{scope} 
	\node [state] (m11) [label=center:{\(m_1\)}] {};
	\node [state] (m21) [above of=m11,label=left:{\(\SerAmp{a}\)},label=center:{\(m_2\)}] {}
	edge 	[pre] (m11);
	\node [state] (m31) [above of=m21,label=center:{\(m_3\)}] {}
	edge 	[pre] 	(m21);

	\node [state] (m12) [right=0.2cm of m11,label=center:{\(m_1\)}] {};
	\node [state] (m22) [above of=m12,label=right:{\(\SerAmp{a'}\)},label=center:{\(m_2'\)}] {}
	edge 	[pre]	(m12);
	\node [state] (m32) [above of=m22,label=center:{\(m_3\)}] {}
	edge 	[pre]  (m22);
  \end{scope}

 \begin{scope}[xshift=2.7cm] 

	\node [state] (m'12) [label=center:{\(m_1\)}] {};
	\node [nostate] (m'22) [above of=m'12] {};

	\node [state] (m'21) [left=-0.2cm of m'22,label=left:{\(\SerAmp{a}\)},label=center:{\(m_2\)}] {}
	edge [pre] (m'12);
\node [state] (m'23) [right=-0.2cm of m'22,label=right:{\(\SerAmp{a'}\)},label=center:{\(m_2'\)}] {}
	edge [pre] (m'12);

	\node [state] (m'32) [above of=m'22,label=above:{\(\SerAmp{c}\)},label=center:{\(m_3\)}] {}
	edge [pre] (m'21)
	edge [pre] (m'23);

  \end{scope}

  \begin{scope}[xshift=5cm] 
	\node [state] (m1) [label=center:{\(m_1\)}] {};
	\node [elstate] (m2) [above of=m1,label=right:{\(\SerAmp{c}=\Sum{\SerAmp{a}}{\SerAmp{a'}}\)}] {\(m_2\cup m_2'\)}
	edge 	[pre]	(m1);
	\node [state] (m3) [above of=m2,label=center:{\(m_3\)}] {}
	edge 	[pre] 	(m2);
  \end{scope}

 \draw [->,thick,line width=.4mm]
    ([xshift=0.7cm]m22 -| m22) -- ([xshift=-0.7cm]m'21 -| m'21);

 \draw [->,thick,line width=.4mm]
    ([xshift=0.7cm]m'23 -| m'23) -- ([xshift=-0.8cm]m2 -| m2);

\TikZend
  \caption{\emph{Sum Rule}: \(\SerAmp{a}\) and \(\SerAmp{a'}\) are the amplitudes of the two sequences. The amplitude of their coarse graining is \(\SerAmp{c}=\Sum{\SerAmp{a}}{\SerAmp{b}}\).}
  \label{fig:coarse}
\end{figure}

Another operation is coarse-graining: given two sequences of the same length, differing only by a single, intermediate outcome, we can posit a measurement which would not distinguish those outcomes. For example, in Stern-Gerlach experiments, this could correspond to simply detecting whether an atom passed through the apparatus, regardless of its spin. Similarly, in a multiple-slit experiment, this could be an outcome of an experiment in which the slit detectors are inaccurate, so they would group several slits into a single outcome. Then, we may combine these results in parallel, and the amplitude for the new series is the sum of those for the old. This is the \emph{Sum Rule} (see Fig.~\ref{fig:coarse}).

Finally, for measurements that immediately follow each other in time, the 
amplitude for, say, \(m_1\) followed immediately by \(m_2\) is the complex conjugate of 
the amplitude for \(m_2\) followed immediately by \(m_1\). This is called \emph{Reciprocity} (See Fig.~\ref{fig:inv}).

\begin{figure}[!ht] 
\centering
\TikZstartFeynmanSmall

  \begin{scope}
	\node [state] (m11) [label=center:{\(m_1\)}] {};
	\node [state] (m21) [right of=m11,label=center:{\(m_2\)}] {}
	edge 	[pre] node[above] {\(\SerAmp{a}\)} (m11);
	\node [nostate] (arr) [right=0.5cm of m21] {\(\Rightarrow\)};
	\node [state] (m12) [right=0.5cm of arr,label=center:{\(m_1\)}] {};
	\node [state] (m22) [right of=m12,label=center:{\(m_2\)}] {}
	edge [post] node[above] {\(\Cc{\SerAmp{a}}\)} (m12);

  \end{scope}

\TikZend
  \caption{\emph{Reciprocity}: \(\SerAmp{a}\) is the amplitude of a process with two measurements, one immediately after the other. The amplitude when the measurements are interchanged is \(\Cc{\SerAmp{a}}\).}
  \label{fig:inv}
\end{figure}
\section{Two Particles}

\begin{figure}[!ht]
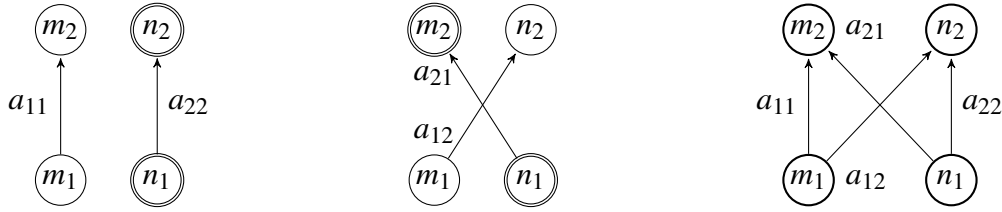
 
\centering
\TikZstartFeynmanMedium

  \begin{scope} 
	\node [astate] (m11) [label=center:{\(m_1\)}] {};
	\node [astate] (m21) [above of=m11,label=center:{\(m_2\)}] {}
	edge [pre] node[left] {\(\AmpEl{a}{1}{1}\)} (m11);

	\node [bstate] (m12) [right=0.6cm of m11,label=center:{\(n_1\)}] {};
	\node [bstate] (m22) [above of=m12,label=center:{\(n_2\)}] {}
	edge [pre] node[right] {\(\AmpEl{a}{2}{2}\)} (m12);
  \end{scope}

  \begin{scope} 
	\node [astate] (m'11) [right=3.0cm of m12,label=center:{\(m_1\)},label=above:{\(\AmpEl{a}{1}{2}\)}] {};
	\node [bstate] (m'12) [right=0.6cm of m'11,label=center:{\(n_1\)}] {};
	\node [bstate] (m'21) [above of=m'11,label=center:{\(m_2\)},label=below:{\(\AmpEl{a}{2}{1}\)}] {}
	edge [pre] (m'12);
	\node [astate] (m'22) [above of=m'12,label=center:{\(n_2\)}] {}
	edge [pre] (m'11);
  \end{scope}

  \begin{scope} 
	\node [state] (m''11) [right=3.0cm of m'12,label=center:{\(m_1\)},label=right:{\(\AmpEl{a}{1}{2}\)}] {};
	\node [state] (m''12) [right=1.2cm of m''11,label=center:{\(n_1\)}] {};
	\node [state] (m''21) [above of=m''11,label=center:{\(m_2\)},label=right:{\(\AmpEl{a}{2}{1}\)}] {}
	edge [pre] node[left] {\(\AmpEl{a}{1}{1}\)} (m''11)
	edge [pre] (m''12);
	\node [state] (m''22) [above of=m''12,label=center:{\(n_2\)}] {}
	edge [pre] node[right] {\(\AmpEl{a}{2}{2}\)} (m''12)
	edge [pre] (m''11);
  \end{scope}

\TikZend
  \caption{\emph{Identical Particle Ambiguity}: Had we been able to distinguish the particles, we could label their states differently (single or double thin circles) and treat their transitions separately. This is impossible, so we must incorporate transitions between all states, which we represent as thick circles.}
  \label{fig:twopartamb}
\end{figure}

Consider measurements on a system of two particles.
Let us assume that in the first measurement, we detect two particles, one with outcome \(m_1\) and one with \(n_1\), while in the second measurement, we find one in \(m_2\) and one in \(n_2\). If we knew that the particle in \(m_1\) went to \(m_2\), and the one in \(n_1\) to 
\(n_2\), we would only worry about the direct amplitudes: \(\AmpEl{a}{1}{1}\) and \(\AmpEl{a}{2}{2}\). But if the particles are identical, we cannot distinguish this from a particle going from \(m_1\) to \(n_2\), and the other from \(n_1\) to \(m_2\). In fact, these distinctions are meaningless. Therefore, we have to worry about all four transition amplitudes. (See Fig.~\ref{fig:twopartamb}.)

\begin{figure}[!ht]
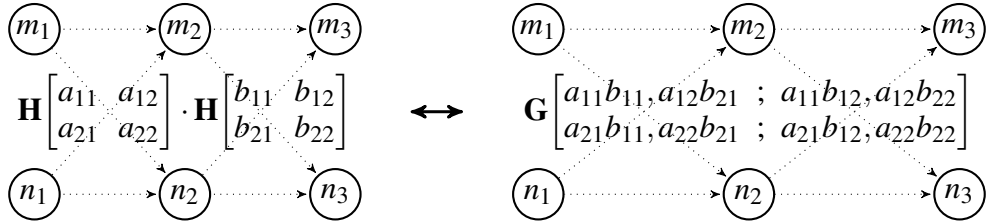
 
\centering
\TikZstartFeynmanReg

  \begin{scope}[node distance=2cm] 
	\node [state] (m11) [label=center:{\(m_1\)}] {};
	\node [state] (m21) [right of=m11,label=center:{\(m_2\)}] {}
	edge [pre,dotted]  (m11);
	\node [state] (m31) [right of=m21,label=center:{\(m_3\)}] {}
	edge [pre,dotted]  (m21);

	\node [nostate] (m12) [below=0.4cm of m11] {};
	\node [nostate] (m22) [right of=m12,label=center:{\(\Product{\Htwo{\AmpEl{a}{1}{1}}{\AmpEl{a}{1}{2}}{\AmpEl{a}{2}{1}}{\AmpEl{a}{2}{2}}}{\Htwo{\AmpEl{b}{1}{1}}{\AmpEl{b}{1}{2}}{\AmpEl{b}{2}{1}}{\AmpEl{b}{2}{2}}}\)}] {};
	\node [nostate] (m32) [right of=m22] {};

	\node [state] (m13) [below=0.4cm of m12,label=center:{\(n_1\)}] {}
	edge [post,dotted] (m21);
	\node [state] (m23) [right of=m13,label=center:{\(n_2\)}] {}
	edge [post, dotted] (m31)
	edge [pre,dotted] (m11)
	edge [pre,dotted] (m13);
	\node [state] (m33) [right of=m23,label=center:{\(n_3\)}] {}
	edge [pre,dotted] (m21)
	edge [pre,dotted] (m23);
  \end{scope}

  \begin{scope} 
	\node [state] (m'11) [right=2cm of m31,label=center:{\(m_1\)}] {};
	\node [state] (m'21) [right of=m'11,label=center:{\(m_2\)}] {}
	edge [pre,dotted]  (m'11);
	\node [state] (m'31) [right of=m'21,label=center:{\(m_3\)}] {}
	edge [pre,dotted]  (m'21);

	\node [nostate] (m'12) [below=0.4cm of m'11] {};
	\node [nostate] (m'22) [right of=m'12,label=center:{\(\Gtwo{\AmpEl{a}{1}{1} \AmpEl{b}{1}{1}, \AmpEl{a}{1}{2} \AmpEl{b}{2}{1}}{\AmpEl{a}{1}{1} \AmpEl{b}{1}{2}, \AmpEl{a}{1}{2} \AmpEl{b}{2}{2}}{\AmpEl{a}{2}{1} \AmpEl{b}{1}{1}, \AmpEl{a}{2}{2} \AmpEl{b}{2}{1}}{\AmpEl{a}{2}{1} \AmpEl{b}{1}{2}, \AmpEl{a}{2}{2} \AmpEl{b}{2}{2}}\)}] {};
	\node [nostate] (m'32) [right of=m'22] {};

	\node [state] (m'13) [below=0.4cm of m'12,label=center:{\(n_1\)}] {}
	edge [post,dotted] (m'21);
	\node [state] (m'23) [right of=m'13,label=center:{\(n_2\)}] {}
	edge [post, dotted] (m'31)
	edge [pre,dotted] (m'11)
	edge [pre,dotted] (m'13);
	\node [state] (m'33) [right of=m'23,label=center:{\(n_3\)}] {}
	edge [pre,dotted] (m'21)
	edge [pre,dotted] (m'23);
  \end{scope}

\draw [<->,thick,line width=.4mm]
    ([xshift=0.5cm]m32 -| m32) -- ([xshift=-0.5cm]m'12 -| m'12);

\TikZend
  \caption{\emph{Product Rule, Twice}: First for the overall amplitude, then for the single-particle amplitudes.}
  \label{fig:concat2id}
\end{figure}

Therefore, our first postulate is that the amplitude for the non-interacting two-particle system is a function of all the relevant single-particle amplitudes, expressed as follows:
\[
\Hop=\Htwo{\AmpEl{a}{1}{1}}{\AmpEl{a}{1}{2}}{\AmpEl{a}{2}{1}}{\AmpEl{a}{2}{2}}\text{where }\Hop,\AmpEl{a}{i}{j}\in\C\text{, and } \abs{\Hop},\abs{\AmpEl{a}{i}{j}}\le1\text{.}
\]
If we have three measurements in series, we can use the product rule, as for any 
system. This means that the total amplitude is a product of two \(\Hop\)'s. But we can also look at all possible paths from start to finish, and use the product rule on each of these paths. We can then reapply the idea that led to our first postulate, and so postulate that the amplitude here is  a function, \(\Gop\), of all 8 relevant single-particle amplitudes. (See Fig.~\ref{fig:concat2id}.) 

Thus we get our first functional equation, connecting \(\Gop\) and \(\Hop\):
\begin{equation}\label{eq:2pconcat}
\Htwo{\AmpEl{a}{1}{1}}{\AmpEl{a}{1}{2}}{\AmpEl{a}{2}{1}}{\AmpEl{a}{2}{2}}\Htwo{\AmpEl{b}{1}{1}}{\AmpEl{b}{1}{2}}{\AmpEl{b}{2}{1}}{\AmpEl{b}{2}{2}} = 
\Gtwo{\AmpEl{a}{1}{1} \AmpEl{b}{1}{1}, \AmpEl{a}{1}{2} \AmpEl{b}{2}{1}}{\AmpEl{a}{1}{1} \AmpEl{b}{1}{2}, \AmpEl{a}{1}{2} \AmpEl{b}{2}{2}}{\AmpEl{a}{2}{1} \AmpEl{b}{1}{1}, \AmpEl{a}{2}{2} \AmpEl{b}{2}{1}}{\AmpEl{a}{2}{1} \AmpEl{b}{1}{2}, \AmpEl{a}{2}{2} \AmpEl{b}{2}{2}}
\text{.}
\end{equation}

Before exploiting it in full, let exhaust the Sum Rule. We choose a simple transition between the last two measurements, giving us the following, simplified equation:
\begin{equation}
\Htwo{\AmpEl{a}{1}{1}}{ \AmpEl{a}{1}{2}}{\AmpEl{a}{2}{1}}{\AmpEl{a}{2}{2}}\Htwo{c}{ 0}{0}{c}=\Gtwo{\AmpEl{a}{1}{1} c, 0}{0, \AmpEl{a}{1}{2} c}{\AmpEl{a}{2}{1} c, 0}{0, \AmpEl{a}{2}{2} c}\text{,}
\end{equation}
\[
\TikZstartFeynmanMedium
 \begin{scope} 
	\node [state] (m12) [label=center:{\(m_1\)}] {};
	\node [state] (m22) [right of=m12,label=center:{\(m_2\)}] {}
	edge [pre,dotted] (m12);
	\node [state] (m32) [right of=m22,label=center:{\(m_3\)}] {}
	edge [pre] (m22);

	\node [state] (m13) [below=0.4cm of m12,label=center:{\(n_1\)}] {}
	edge [post,dotted] (m22);
	\node [state] (m23) [right of=m13,label=center:{\(n_2\)}] {}
	edge [pre,dotted] (m12)
	edge [pre,dotted] (m13);
	\node [state] (m33) [right of=m23,label=center:{\(n_3\)}] {}
	edge [pre] (m23);
  \end{scope}
\TikZend
\]

and coarsen over each middle measurement in turn:
\begin{equation}\label{eq:mcoarsen}
\Htwo{\AmpEl{a}{1}{1}+\AmpEl{a'}{1}{1}}{\AmpEl{a}{1}{2}}{\AmpEl{a}{2}{1}+\AmpEl{a'}{2}{1}}{\AmpEl{a}{2}{2}}\cancelto{1}{\Htwo{c}{0}{0}{c}} = \left(\Htwo{\AmpEl{a}{1}{1}}{ \AmpEl{a}{1}{2}}{\AmpEl{a}{2}{1}}{\AmpEl{a}{2}{2}}+ \Htwo{\AmpEl{a'}{1}{1}}{ \AmpEl{a}{1}{2}}{\AmpEl{a'}{2}{1}}{\AmpEl{a}{2}{2}}\right)\cancelto{1}{\Htwo{c}{0}{0}{c}}
\end{equation}
\[
\TikZstartFeynmanMedium
 \begin{scope} 
	\node [state] (m'12) [label=center:{\(m_1\)}] {};
	\node [elstate] (m'22) [right of=m'12] {\(m_2\cup m_2'\)}
	edge [pre,dotted] (m'12);
	\node [state] (m32) [right of=m'22,label=center:{\(m_3\)}] {}
	edge [pre] (m'22);
	\node [state] (m'13) [below=0.4cm of m'12,label=center:{\(n_1\)}] {}
	edge [post,dotted] (m'22);
	\node [state] (m'23) [right of=m'13,label=center:{\(n_2\)}] {}
	edge [pre,dotted] (m'12)
	edge [pre,dotted] (m'13);
	\node [state] (m33) [right of=m'23,label=center:{\(n_3\)}] {}
	edge [pre] (m'23);
\end{scope}
\TikZend
\]
\begin{equation}\label{eq:ncoarsen}
\Htwo{\AmpEl{a}{1}{1}}{\AmpEl{a}{1}{2}+\AmpEl{a'}{1}{2}}{\AmpEl{a}{2}{1}}{\AmpEl{a}{2}{2}+\AmpEl{a'}{2}{2}}\cancelto{1}{\Htwo{c}{0}{0}{c}} = \left(\Htwo{\AmpEl{a}{1}{1}}{ \AmpEl{a}{1}{2}}{\AmpEl{a}{2}{1}}{\AmpEl{a}{2}{2}}+ \Htwo{\AmpEl{a}{1}{1}}{ \AmpEl{a'}{1}{2}}{\AmpEl{a}{2}{1}}{\AmpEl{a'}{2}{2}}\right)\cancelto{1}{\Htwo{c}{0}{0}{c}}
\end{equation}
\[
\TikZstartFeynmanMedium
 \begin{scope} 
	\node [state] (m''12) [label=center:{\(m_1\)}] {};
	\node [state] (m''22) [right of=m''12,label=center:{\(m_2\)}] {}
	edge [pre,dotted] (m''12);
	\node [state] (m32) [right of=m''22,label=center:{\(m_3\)}] {}
	edge [pre] (m''22);

	\node [state] (m''13) [below=0.4cm of m''12,label=center:{\(n_1\)}] {}
	edge [post,dotted] (m''22);
	\node [elstate] (m''23) [right of=m''13] {\(n_2\cup n_2'\)}
	edge [pre,dotted] (m''12)
	edge [pre,dotted] (m''13);
	\node [state] (m33) [right of=m''23,label=center:{\(n_3\)}] {}
	edge [pre] (m''23);
  \end{scope}
\TikZend
\] 

We see that \(\Htwosmall{c}{0}{0}{c}\) factors out, and we get two additivity conditions. Eq.~\eqref{eq:mcoarsen} gives us:
\begin{equation}
\Htwo{\AmpEl{a}{1}{1}}{\AmpEl{a}{1}{2}}{\AmpEl{a}{2}{1}}{\AmpEl{a}{2}{2}}=\Htwo{\AmpEl{a}{1}{1}+0}{\AmpEl{a}{1}{2}}{0+\AmpEl{a}{2}{1}}{\AmpEl{a}{2}{2}} = 
\Htwo{\AmpEl{a}{1}{1}}{ \AmpEl{a}{1}{2}}{0}{\AmpEl{a}{2}{2}}+\Htwo{0}{\AmpEl{a}{1}{2}}{\AmpEl{a}{2}{1}}{\AmpEl{a}{2}{2}}.
\end{equation}
We can then apply Eq.~\eqref{eq:ncoarsen} to each of these terms to get:
\begin{equation}
\Htwo{\AmpEl{a}{1}{1}}{ \AmpEl{a}{1}{2}}{\AmpEl{a}{2}{1}}{\AmpEl{a}{2}{2}}=\Htwo{\AmpEl{a}{1}{1}}{0}{0}{\AmpEl{a}{2}{2}}+\cancelto{0}{\Htwo{\AmpEl{a}{1}{1}}{\AmpEl{a}{1}{2}}{0}{0}}+
\cancelto{0}{\Htwo{0}{0}{\AmpEl{a}{2}{1}}{\AmpEl{a}{2}{2}}}+\Htwo{0}{\AmpEl{a}{1}{2}}{\AmpEl{a}{2}{1}}{0}.
\end{equation}
Two of these four terms are inconsistent with having two particles in two measurements.

\begin{figure}[!ht]
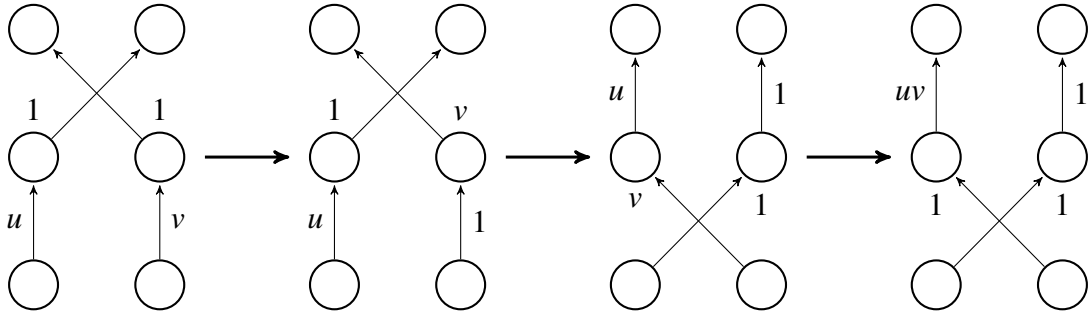
 
\centering
\TikZstartFeynmanSmall

 \begin{scope} 
	\node [state] (m11) {};
	\node [state] (m21) [above of=m11,label=above:{\(1\)}] {}
	edge [pre] node[left] {\(\SerAmp{u}\)} (m11);
	\node [state] (m31) [above of=m21] {};

	\node [state] (m12) [right=1cm of m11] {};
	\node [state] (m22) [above of=m12,label=above:{\(1\)}] {}
	edge [pre] node[right] {\(\SerAmp{v}\)} (m12)
	edge [post] (m31);
	\node [state] (m32) [above of=m22] {}
	edge [pre] (m21);
  \end{scope}  

 \begin{scope}[xshift=2cm] 
	\node [state] (m'11) {};
	\node [state] (m'21) [above of=m'11,label=above:{\(1\)}] {}
	edge [pre] node[left] {\(\SerAmp{u}\)} (m'11);
	\node [state] (m'31) [above of=m'21] {};

	\node [state] (m'12) [right=1cm of m'11] {};
	\node [state] (m'22) [above of=m'12,label=above:{\(\SerAmp{v}\)}] {}
	edge [pre] node[right] {\(1\)} (m'12)
	edge [post] (m'31);
	\node [state] (m'32) [above of=m'22] {}
	edge [pre] (m'21);
  \end{scope}  

\begin{scope}[xshift=4cm] 
	\node [state] (m''11) {};
	\node [state] (m''21) [above of=m''11,label=below:{\(\SerAmp{v}\)}] {};
	\node [state] (m''31) [above of=m''21] {}
	edge [pre] node[left] {\(\SerAmp{u}\)} (m''21);

	\node [state] (m''12) [right=1cm of m''11] {}
	edge [post] (m''21);
	\node [state] (m''22) [above of=m''12,label=below:{\(1\)}] {}
	edge [pre] (m''11);
	\node [state] (m''32) [above of=m''22] {}
	edge [pre] node[right] {\(1\)} (m''22);
  \end{scope}  

\begin{scope}[xshift=6cm] 
	\node [state] (m'''11) {};
	\node [state] (m'''21) [above of=m'''11,label=below:{\(1\)}] {};
	\node [state] (m'''31) [above of=m'''21] {}
	edge [pre] node[left] {\(\SerAmp{uv}\)} (m'''21);

	\node [state] (m'''12) [right=1cm of m'''11] {}
	edge [post] (m'''21);
	\node [state] (m'''22) [above of=m'''12,label=below:{\(1\)}] {}
	edge [pre] (m'''11);
	\node [state] (m'''32) [above of=m'''22] {}
	edge [pre] node[right] {\(1\)} (m'''22);
  \end{scope}  

 \draw [->,thick,line width=.4mm]
    ([xshift=.3cm]m22 -| m22) -- ([xshift=-.3cm]m'21 -| m'21);

 \draw [->,thick,line width=.4mm]
    ([xshift=.3cm]m'22 -| m'22) -- ([xshift=-.3cm]m''21 -| m''21);

\draw [->,thick,line width=.4mm]
    ([xshift=.3cm]m''22 -| m''22) -- ([xshift=-.3cm]m'''21 -| m'''21);

\TikZend
  \caption{\emph{Simplifying the First Term}: Eq.~\eqref{eq:2pconcat} allows us to slide single-particle amplitudes along paths. The overall \(\Hop\) amplitudes can also have their order changed, justifying the second transition.}
  \label{fig:simplifyfirst}
\end{figure}

It is in simplifying the remaining two that Eq.~\eqref{eq:2pconcat} truly shines, allowing us to move single-particle amplitudes along paths. Combining the first term with a trivial transposition as the second process, we can perform a series of 
simplifications  (See Fig.~\ref{fig:simplifyfirst}):
\begin{equation}
\Htwo{\SerAmp{u}}{0}{0}{\SerAmp{v}}\Htwo{0}{1}{1}{0}=\Htwo{\SerAmp{u}}{0}{0}{1}\Htwo{0}{1}{v}{0}=\Htwo{0}{1}{v}{0}\Htwo{\SerAmp{u}}{0}{0}{1}=\Htwo{0}{1}{1}{0}\Htwo{\SerAmp{u}\SerAmp{v}}{0}{0}{1}\text{;}
\end{equation}
thus the first term is a single-parameter function.

\begin{figure}[!ht]
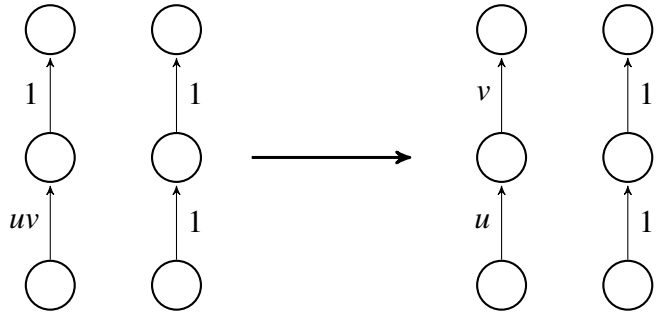
 
\centering
\TikZstartFeynmanSmall

 \begin{scope} 
	\node [state] (m11) {};
	\node [state] (m21) [above of=m11] {}
	edge [pre] node[left] {\(\SerAmp{uv}\)} (m11);
	\node [state] (m31) [above of=m21] {}
	edge [pre] node[left] {\(1\)} (m21);

	\node [state] (m12) [right=1cm of m11] {};
	\node [state] (m22) [above of=m12] {}
	edge [pre] node[right] {1} (m12);
	\node [state] (m32) [above of=m22] {}
	edge [pre] node[right] {1} (m22);
  \end{scope}  

 \begin{scope}[xshift=3cm] 
	\node [state] (m'11) {};
	\node [state] (m'21) [above of=m'11] {}
	edge [pre] node[left] {\(\SerAmp{u}\)} (m'11);
	\node [state] (m'31) [above of=m'21] {}
	edge [pre] node[left] {\(\SerAmp{v}\)} (m'21);

	\node [state] (m'12) [right=1cm of m'11] {};
	\node [state] (m'22) [above of=m'12] {}
	edge [pre] node[right] {1} (m'12);
	\node [state] (m'32) [above of=m'22] {}
	edge [pre] node[right] {1} (m'22);
  \end{scope}  

 \draw [->,thick,line width=.4mm]
    ([xshift=.5cm]m22 -| m22) -- ([xshift=-.6cm]m'21 -| m'21);

\TikZend
  \caption{\emph{Justifying Multiplicativity}: as in Fig.~\ref{fig:simplifyfirst}, one can slide amplitudes along paths.}
  \label{fig:mult}
\end{figure}

A similar combination reduces the second term to the first term up to a constant factor:
\begin{equation}
\Htwo{0}{\SerAmp{u}}{\SerAmp{v}}{0}\Htwo{1}{0}{0}{1} = \Htwo{0}{1}{1}{0}\Htwo{\SerAmp{v}}{0}{0}{\SerAmp{u}}.
\end{equation}
We end up with the following expression, containing only one functional dependence:
\begin{equation}\label{eq:onefuncdep}
\Htwo{\AmpEl{a}{1}{1}}{ \AmpEl{a}{1}{2}}{\AmpEl{a}{2}{1}}{\AmpEl{a}{2}{2}}=\Htwo{\AmpEl{a}{1}{1}\AmpEl{a}{2}{2}}{0}{0}{1}+\left(\Htwo{0}{1}{1}{0}\div\Htwo{1}{0}{0}{1}\right)\Htwo{\AmpEl{a}{1}{2}\AmpEl{a}{2}{1}}{0}{0}{1}\text{.}
\end{equation}
This function satisfies not only additivity (from~Eq.~\eqref{eq:mcoarsen}):
\begin{equation}
\Htwo{\SerAmp{u}+\SerAmp{v}}{0}{0}{1} = \Htwo{\SerAmp{u}}{0}{0}{1}+\Htwo{\SerAmp{v}}{0}{0}{1}\text{,}
\end{equation}
but also a form of multiplicativity (See Fig.~\ref{fig:mult}):
\begin{equation}
\Htwo{\SerAmp{u}\SerAmp{v}}{0}{0}{1}\Htwo{1}{0}{0}{1} = \Htwo{\SerAmp{u}}{0}{0}{1}\Htwo{\SerAmp{v}}{0}{0}{1}\text{.}
\end{equation}

If we now define a new function:
\begin{equation}
\f(z)\triangleq \Htwo{z}{0}{0}{1}\div\Htwo{1}{0}{0}{1}
\end{equation}
to get the following pair of complex functional equations in complex variables:
\begin{align}
	\f(\SerAmp{u}\SerAmp{v}) &= \f(\SerAmp{u})\f(\SerAmp{v})\\
	\f(\SerAmp{u}+\SerAmp{v}) &= \f(\SerAmp{u})+\f(\SerAmp{v})\text{, where \(u, v\in\C\).}
\end{align}

This restrictive pair of equations allows only two non-trivial, continuous solutions:
\begin{equation}
\f(z) = \left\{
			\begin{matrix}
				z&\text{(\emph{the identity})}\\
				\Cc{z}&\text{(\emph{the complex conjugate})}\\
			\end{matrix}
		\right.
\end{equation}

 If we rewrite the amplitude from Eq.~\eqref{eq:onefuncdep} in terms of \(\f\), we get the following:
\begin{equation}
\Htwo{\AmpEl{a}{1}{1}}{ \AmpEl{a}{1}{2}}{\AmpEl{a}{2}{1}}{\AmpEl{a}{2}{2}} = \Htwo{1}{0}{0}{1}\f(\AmpEl{a}{1}{1}\AmpEl{a}{2}{2})+\Htwo{0}{1}{1}{0}\f(\AmpEl{a}{1}{2}\AmpEl{a}{2}{1})\text{.}
\end{equation}

This leaves us with two complex constants; fortunately, they, too, may be restricted. Recall \emph{Reciprocity} (Fig.~\ref{fig:inv}). Reversing the order of the two-particle measurements reverses the order of each of the individual one-particle measurements. We can represent this either by taking the complex conjugate of the total amplitude, \(\Hop\), or by applying \(\Hop\) to the complex conjugates of the single-particle amplitudes, \(\AmpEl{a}{i}{j}\), leading to the identity:
\begin{equation}
\Cc{\Htwo{\AmpEl{a}{1}{1}}{\AmpEl{a}{1}{2}}{\AmpEl{a}{2}{1}}{\AmpEl{a}{2}{2}}}=\Htwo{\Cc{\AmpEl{a}{1}{1}}}{\Cc{\AmpEl{a}{1}{2}}}{\Cc{\AmpEl{a}{2}{1}}}{\Cc{\AmpEl{a}{2}{2}}}.
\end{equation}
When the single-particle amplitudes are real, so are the two-particle amplitudes:
\begin{align}
\Cc{\Htwo{1}{0}{0}{1}}= \Htwo{\Cc{1}}{\Cc{0}}{\Cc{0}}{\Cc{1}}=\Htwo{1}{0}{0}{1}\text{, }
\Cc{\Htwo{0}{1}{1}{0}}= \Htwo{\Cc{0}}{\Cc{1}}{\Cc{1}}{\Cc{0}}=\Htwo{0}{1}{1}{0},
\end{align}
so that \(\Htwosmall{1}{0}{0}{1},\Htwosmall{0}{1}{1}{0}\in\R\). Let us now choose a special case. The single-particle amplitudes in \(\Htwosmall{1}{0}{0}{1}\) could correspond to a system in which tunnels connect outcome \(m_1\) to~\(m_2\), and~\(n_1\) to~\(n_2\), respectively. The tunnels could then be very far removed in space. The particles being identical does not change the fact that both single-particle processes are deterministic, and therefore, that the process for the system as a whole is deterministic~(\(\Pr=1\)). A similar argument works for \(\Htwosmall{0}{1}{1}{0}\) (think of the English Channel). This special case must generalize, as \(\Hop\) is a function of the amplitudes \(\AmpEl{a}{i}{j}\), regardless of the choice of experiment. Therefore:
\begin{align}
\left|\Htwo{1}{0}{0}{1}\right|^2=\left|\Htwo{0}{1}{1}{0}\right|^2=1\xRightarrow{(\textit{both real})}\Htwo{1}{0}{0}{1}=\pm1\text{, }\Htwo{0}{1}{1}{0}=\pm1.
\end{align}

The general solution for two particles is then:
\begin{equation}
\Htwo{\AmpEl{a}{1}{1}}{ \AmpEl{a}{1}{2}}{\AmpEl{a}{2}{1}}{\AmpEl{a}{2}{2}} = \left\{
			\begin{matrix}
				\pm\left(\AmpEl{a}{1}{1}\AmpEl{a}{2}{2}\pm\AmpEl{a}{1}{2}\AmpEl{a}{2}{1}\right)\text{, \emph{with the two \(\pm\) independent}\phantom{.}}\\[10pt]
				\Cc{\pm\left(\AmpEl{a}{1}{1}\AmpEl{a}{2}{2}\pm\AmpEl{a}{1}{2}\AmpEl{a}{2}{1}\right)}\text{, \emph{with the two \(\pm\) independent}.}\\
			\end{matrix}
		\right.
\end{equation}

Complex conjugation distributes over addition and multiplication, so if we choose any of the lower possibilities, the total amplitude will always be the complex conjugate of the total amplitude from choosing the corresponding upper possibility. Similarly, two choices separated by a minus sign will always lead to the same total amplitude up to \(\pm\). Neither of these will alter the modulus squared of the amplitude, which is what leads to physical predictions. This allows us to regraduate, i.e., to choose, for convenience, that first term will have a \(+\) sign, and the single-particle amplitudes appear rather than their complex conjugates, showing, as expected, that there are only two physically distinguishable alternatives:
\begin{equation}
\Htwo{\AmpEl{a}{1}{1}}{ \AmpEl{a}{1}{2}}{\AmpEl{a}{2}{1}}{\AmpEl{a}{2}{2}} = \left\{
			\begin{matrix}
				\AmpEl{a}{1}{1}\AmpEl{a}{2}{2}+\AmpEl{a}{1}{2}\AmpEl{a}{2}{1}&\text{(\emph{bosons})\phantom{.}}\\
				\AmpEl{a}{1}{1}\AmpEl{a}{2}{2}-\AmpEl{a}{1}{2}\AmpEl{a}{2}{1}&\text{(\emph{fermions}).}\\
			\end{matrix}
		\right.
\end{equation}

The reader will note that clarifying the subject matter and improving the notation over~\cite{Tikochinsky1988} has allowed us to use weaker assumptions than analyticity to prove this result, and, in fact, that we have used the Feynman rules much more extensively and fruitfully.
\section{Three or More Particles}
The methods employed so far easily extend to 3 or more particles, as we shall briefly outline here. The generalization to \(N\) interacting particles will be found in Ref.~\cite{GoyalNeori2013Unp}.

The sum rule will give us the sum of all terms corresponding to 
appropriate transitions, in this case those with three particles at each 
measurement:
\begin{align}
\Hthree{\AmpEl{a}{1}{1}}{\AmpEl{a}{1}{2}}{\AmpEl{a}{1}{3}}{\AmpEl{a}{2}{1}}{\AmpEl{a}{2}{2}}{\AmpEl{a}{2}{3}}{\AmpEl{a}{3}{1}}{\AmpEl{a}{3}{2}}{\AmpEl{a}{3}{3}} &=
\Hthree{\AmpEl{a}{1}{1}}{0}{0}{0}{\AmpEl{a}{2}{2}}{0}{0}{0}{\AmpEl{a}{3}{3}}+\Hthree{0}{\AmpEl{a}{1}{2}}{0}{0}{0}{\AmpEl{a}{2}{3}}{\AmpEl{a}{3}{1}}{0}{0}+\Hthree{0}{0}{\AmpEl{a}{1}{3}}{\AmpEl{a}{2}{1}}{0}{0}{0}{\AmpEl{a}{3}{2}}{0}+ \nonumber \\
&+\Hthree{\AmpEl{a}{1}{1}}{0}{0}{0}{0}{\AmpEl{a}{2}{3}}{0}{\AmpEl{a}{3}{2}}{0}+\Hthree{0}{\AmpEl{a}{1}{2}}{0}{\AmpEl{a}{2}{1}}{0}{0}{0}{0}{\AmpEl{a}{3}{3}}+\Hthree{0}{0}{\AmpEl{a}{1}{3}}{0}{\AmpEl{a}{2}{2}}{0}{\AmpEl{a}{3}{1}}{0}{0}.
\end{align}

Iterating and reordering the product rules, followed regraduation, will lead to:
\begin{align}
\Hthree{\AmpEl{a}{1}{1}}{\AmpEl{a}{1}{2}}{\AmpEl{a}{1}{3}}{\AmpEl{a}{2}{1}}{\AmpEl{a}{2}{2}}{\AmpEl{a}{2}{3}}{\AmpEl{a}{3}{1}}{\AmpEl{a}{3}{2}}{\AmpEl{a}{3}{3}} &=
\AmpEl{a}{1}{1}\AmpEl{a}{2}{2}\AmpEl{a}{3}{3}\Hthreesmall{1}{0}{0}{0}{1}{0}{0}{0}{1}+\AmpEl{a}{1}{2}\AmpEl{a}{2}{3}\AmpEl{a}{3}{1}\Hthreesmall{0}{1}{0}{0}{0}{1}{1}{0}{0}+\AmpEl{a}{1}{3}\AmpEl{a}{2}{1}\AmpEl{a}{3}{2}\Hthreesmall{0}{0}{1}{1}{0}{0}{0}{1}{0}+ \nonumber \\
&+\AmpEl{a}{1}{1}\AmpEl{a}{2}{3}\AmpEl{a}{3}{2}\Hthreesmall{1}{0}{0}{0}{0}{1}{0}{1}{0} +\AmpEl{a}{1}{2}\AmpEl{a}{2}{1}\AmpEl{a}{3}{3}\Hthreesmall{0}{1}{0}{1}{0}{0}{0}{0}{1}+\AmpEl{a}{1}{3}\AmpEl{a}{2}{2}\AmpEl{a}{3}{1}\Hthreesmall{0}{0}{1}{0}{1}{0}{1}{0}{0}
\text{,}
\end{align}
where we shall choose \(\Hthreesmall{1}{0}{0}{0}{1}{0}{0}{0}{1}=1\), and the other five \(\Hop\) constants can only be \(\pm1\).

We can reduce the apparent \(2^5=32\) possibilities to two as follows: consider two special cases, each relating only a pair among the six coefficients, twinned via transposition:
\begin{align}
\Hthree{\SerAmp{c}}{0}{0}{0}{\SerAmp{c}}{\SerAmp{c}}{0}{\SerAmp{c}}{\SerAmp{c}}&=\SerAmp{c}^3\left(\Hthreesmall{1}{0}{0}{0}{1}{0}{0}{0}{1}
+ \Hthreesmall{1}{0}{0}{0}{0}{1}{0}{1}{0}\right)\\
\Hthree{0}{c}{0}{c}{0}{c}{c}{0}{c} &=c^3\left(\Hthreesmall{0}{1}{0}{0}{0}{1}{1}{0}{0}
+ \Hthreesmall{0}{1}{0}{1}{0}{0}{0}{0}{1}\right)
\end{align}

Although they may look different at first, note that the labels chosen for the
particles at each measurement are arbitrary, because they are identical. Therefore, 
we can switch labels between the first and second particle without changing the experimental outcome. That means that the probability of both processes must be the same:
\begin{equation}
	\abs{\Hthreesmall{1}{0}{0}{0}{1}{0}{0}{0}{1} + \Hthreesmall{1}{0}{0}{0}{0}{1}{0}{1}{0}}^2 = \abs{\Hthreesmall{0}{1}{0}{0}{0}{1}{1}{0}{0}+ \Hthreesmall{0}{1}{0}{1}{0}{0}{0}{0}{1}}^2
\end{equation}

Which leads us to the end result: since all coefficients are \(\pm1\), the sign must either be \emph{the same} within each pair, or \emph{opposite}. If it is \emph{the same}, all coefficients have \emph{the same} sign, and it is a symmetric expression, corresponding to \emph{bosons}. If it is \emph{opposite}, then all coefficients connected by a transposition are of \emph{opposite} sign, and the expression is anti-symmetric, corresponding to \emph{fermions}.
This proves the symmetrization postulate for three non-interacting particles. This will be generalized to \(N\) \emph{interacting} particles in~\cite{GoyalNeori2013Unp}.

\section{Conclusion}
Let us review our results:
\begin{itemize}
\item We have incorporated identical particles into the Feynman Rules.
\item We have proven the \emph{Symmetrization Postulate} explicitly for two non-interacting particles, and indicated the generalization to \(N\) (to be fully provided in~\cite{GoyalNeori2013Unp}).
\item We have substantially improved upon \cite{Tikochinsky1988} by being more explicit in our formalism, and by introducing weaker assumptions than his implicitly assumed analyticity.
\end{itemize}
\begin{theacknowledgments}
We would like to thank John~Skilling for a very thorough review of our draft, and Ariel~Caticha for insightful discussions, and for suggesting his own operational work~\cite{Caticha2000}.
\end{theacknowledgments}
\bibliographystyle{aipproc} 
\bibliography{KlilMaxEntProcPaper}
\end{document}